# Forthcoming mutual events of planets and astrometric radio sources

## Z. Malkin, V. L'vov, S. Tsekmejster

Pulkovo Observatory, St. Petersburg, Russia

**Abstract.** Radio astronomy observations of close approaches of the Solar system planets to compact radio sources as well as radio source occultations by planets may be of large interest for planetary sciences, dynamical astronomy, and testing gravity theories. In this paper, we present extended lists of occultations of astrometric radio sources observed in the framework of various astrometric and geodetic VLBI programs by planets, and close approaches of planets to radio sources expected in the nearest years. Computations are made making use of the EPOS software package.

**Introduction**

Very long baseline interferometry (VLBI) and single dish radio observations of close approaches of the Solar system bodies (planets, satellites, asteroids) to compact radio sources, as well as radio source occultations by planets may be of large interest for planetary sciences, dynamical astronomy, and testing gravity theories. In this paper we present a new extended version of the list of occultations of astrometric radio sources (i.e. having reliable coordinates at the milliarcsecond level of accuracy) by planets, and close approaches of planets to radio sources expected in the nearest 40 years compiled at the Pulkovo Observatory. Previous lists presented in [1, 2] were substantially revised in two respects. First, a new version of software to find occultation and approaches was used. Second, the list of astrometric radio sources was extended.

All computations of the mutual events circumstances have been made at the Pulkovo Observatory making use of the EPOS software package (http://www.gao.spb.ru/personal/neo/ENG/ESUPP/main.htm) and other programs. The list of astrometric radio sources has been taken from the OCARS catalog available at http://www.gao.spb.ru/english/as/ac_vlbi/ocars.txt.

**Occultations of astrometric radio sources by planets**

Observations of occultations of compact radio sources by Solar system planets may be interesting for several astronomical and physical applications, such as testing GR [1], improvement of planet orbits and their tie to the International Celestial Reference Frame (ICRF) [3], and planetary researches [4, 5]. List of nearest expected events is shown in Table 1 along with the elongation from the Sun. The latter is important for planning of radio astronomy observations. If the radio source is too close to the Sun, it may be impossible to observe it. Nevertheless, we include all the events found during our calculations for completeness.

Occultations of radio sources by planets, like solar eclipses, generally speaking can only be observed in a limited region. A map of the shadow path and detailed circumstances for each VLBI station situated on the shadow path for several selected occultations that will occur in regions with several stations and hence most interesting for radio astronomy experiments are given in [2].

One can see that occultations of astrometric radio sources by planets, especially by outer planets, suitable for radio observations are rare events. However, some physical applications do not require knowledge of accurate source coordinates. For such studies, any compact radio source can be observed, which makes experiment scheduling much easier.

**Close angular approaches of planet to astrometric radio sources**

During close angular approaches of Solar system planets to astrometric radio sources, the apparent positions of these sources shift due to relativistic effects. Thus, these events may be used for testing gravity theories, see, e.g., [1] and paper cited therein. This fact was successfully demonstrated in the experiments on the measurements of radio source position shifts during the approaches of Jupiter carried out in 1988 and 2002 [6, 7]. Basic circumstances for the coming events (10 years in advance for Jupiter and 15 years for other planets) are shown in Table 2.

Table 1. Occultations of astrometric radio sources by planets

| Planet | Date, UT Y M D | Source | Elongation, deg |
|---|---|---|---|
| Venus | 2012 12 24.4 | 1631–208 | 23W |
| Mercury | 2014 07 30.2 | 0750+218 | 11W |
| Venus | 2015 08 06.8 | 0947+064 | 15E |
| Jupiter | 2016 04 10.4 | 1101+077 | 144E |
| Venus | 2020 01 16.7 | 2220–119 | 38E |
| Venus | 2020 07 17.7 | 0446+178 | 42W |
| Mercury | 2022 11 14.7 | 1529–195 | 4E |
| Jupiter | 2025 09 18.6 | 0725+219 | 65W |
| Mercury | 2027 03 21.7 | 2220–119 | 27W |
| Saturn | 2028 10 24.8 | 0223+113 | 173W |
| Mercury | 2029 01 14.3 | 1958–179 | 5E |
| Venus | 2029 02 28.2 | 2221–116 | 6W |
| Mercury | 2029 04 16.1 | 0243+181 | 19E |
| Mercury | 2029 12 27.9 | 1858–212 | 8E |
| Mercury | 2030 02 27.6 | 2208–137 | 9W |
| Jupiter | 2033 02 04.2 | 2104–173 | 1W |

**Summary**

In this paper, we present new lists of found mutual events of astrometric extragalactic radio sources and Solar system planets. The circumstances of both occultations and close angular approaches have been calculated. For these computations, we considered only astrometric radio sources having reliable coordinates at the milliarcsecond level of accuracy. Observations of these sources are used when the tie to the ICRF is important. For other, mainly physical, studies any compact source may be suitable for observations, see, e.g., [3].

For various tasks, VLBI or single-dish radio observations can be performed. Different observing techniques and strategies may require pre-computation of different circumstances of the event of interest. These data can be calculated as well on request using the EPOS software and associated programs developed at the Pulkovo Observatory.

Tables presented above contain only basic circumstances for events expected in limited period of time. Full version of these Tables for all the events found for the period till 2050 with more details is available at the Pulkovo Observatory Web site http://www.gao.spb.ru/english/as/ac_vlbi/#Approaches. Besides, detailed computation of the circumstances for selected events of special interest, including also small Solar system bodies and extended lists of radio sources, can be performed on request.

Table 2. Close angular approaches of planets to astrometric radio sources. In this table, $d$ is the angular distance between the planet and radio source, $E$ is the elongation from the Sun

| Planet | Date, UT Y M D | Source | $d$, arcsec | $E$, deg |
|---|---|---|---|---|
| Jupiter | 2011 07 03.6 | 0210+119 | 341 | 66W |
| | 2011 08 16.8 | 0229+131 | 488 | 104W |
| | 2011 09 13.1 | 0229+131 | 149 | 130W |
| | 2011 11 27.2 | 0156+105 | 285 | 147E |
| | 2012 02 04.0 | 0201+113 | 490 | 78E |
| | 2012 02 20.3 | 0210+119 | 342 | 64E |
| | 2012 04 22.5 | 0300+162 | 115 | 16E |
| | 2013 02 28.1 | 0420+210 | 216 | 88E |
| | 2013 03 29.5 | 0435+217 | 563 | 63E |
| | 2013 10 23.0 | 0723+219 | 123 | 100W |
| | 2013 11 07.2 | 0725+219 | 388 | 114W |
| | 2013 11 22.1 | 0723+219 | 351 | 130W |
| | 2014 07 26.1 | 0814+201 | 488 | 1W |
| | 2014 08 22.8 | 0839+187 | 360 | 21W |
| | 2014 09 09.3 | 0854+178 | 310 | 35W |
| | 2016 10 19.2 | 1229–021 | 506 | 18W |
| | 2017 10 13.7 | 1352–104 | 69 | 10E |
| | 2019 10 20.5 | 1717–229 | 222 | 55E |
| | 2019 10 28.4 | 1723–229 | 184 | 48E |
| | 2020 01 30.3 | 1853–226 | 542 | 27W |
| | 2020 02 15.0 | 1907–224 | 91 | 39W |
| | 2020 08 02.0 | 1922–224 | 78 | 160E |
| | 2020 10 24.2 | 1922–224 | 355 | 79E |
| | 2021 02 19.9 | 2104–173 | 149 | 17W |
| | 2021 03 16.0 | 2126–158 | 528 | 36W |
| | 2021 11 29.8 | 2147–144 | 79 | 77E |
| Saturn | 2013 12 05.8 | 1459–149 | 525 | 26W |
| | 2014 08 26.3 | 1459–149 | 486 | 75E |
| | 2015 06 19.1 | 1548–177 | 156 | 152E |
| | 2015 11 19.1 | 1614–195 | 64 | 10E |
| | 2017 12 13.3 | 1752–225 | 73 | 8E |
| | 2019 11 16.6 | 1907–224 | 240 | 53E |
| | 2021 08 10.7 | 2044–188 | 20 | 171E |
| | 2021 08 19.0 | 2042–191 | 441 | 163E |
| | 2021 12 01.0 | 2042–191 | 382 | 60E |
| | 2021 12 08.1 | 2044–188 | 114 | 53E |
| | 2022 03 11.2 | 2126–158 | 521 | 31W |
| | 2022 05 29.1 | 2147–144 | 288 | 103W |
| | 2023 04 13.3 | 2221–116 | 33 | 49W |
| | 2023 04 18.2 | 2223–114 | 276 | 54W |
| | 2024 01 04.6 | 0220–119 | 370 | 50E |
| | 2024 03 18.5 | 2252–090 | 158 | 16W |
| | 2024 03 28.0 | 2256–084 | 388 | 25W |
| Uranus | 2013 05 05.4 | 0036+030 | 558 | 35W |
| | 2013 10 03.3 | 0036+030 | 362 | 179W |
| | 2016 07 11.8 | 0127+084 | 313 | 86W |
| | 2016 08 16.9 | 0127+084 | 259 | 120W |
| | 2017 04 12.8 | 0127+084 | 499 | 2E |
| Neptune | 2024 09 02.7 | 2354–021 | 498 | 162W |